\documentclass[12pt]{article}
\usepackage{graphicx}
\usepackage{epsfig}
\usepackage{amsmath}
\usepackage{color}

\makeatletter
\@addtoreset{equation}{section}

\renewcommand{\thefootnote}{\fnsymbol{footnote}}
\makeatother

\newcommand {\beq}{\begin{eqnarray}}
\newcommand {\eeq}{\end{eqnarray}}

\newcommand{\spone}{0.9}  

\newcommand{\sptwo}{1.4}
\newcommand{\spthree}{2.4}

\newcommand{\singlespace}{\edef\baselinestretch{\spone}\Large\normalsize}

\newcommand{\doublespace}{\edef\baselinestretch{\sptwo}\Large\normalsize}
\newcommand{\threespace}{\edef\baselinestretch{\spthree}\Large\normalsize}

\singlespace
\threespace
\doublespace

\everymath={\displaystyle}
\begin{document}
\doublespace

~
\vspace{0.5cm}

\begin{center}
{\bf {\Large Renormalization Invariants and\\ 
Quark Flavor Mixings}\\
$~$\\
Lu-Xin Liu  
\\
{
\it Department of Physics, \\
Purdue University,\\ 
West Lafayette, IN 47907, USA\\
and \\
National Institute for Theoretical Physics,\\
Department of Physics and Centre for Theoretical Physics,\\ 
University of the Witwatersrand,\\
Wits, 2050, South Africa\\
Luxin.Liu9@gmail.com\\

\vspace{10pt}
}}
\end{center}
\vspace{15pt}
{\bf Abstract.} 
A set of renormalization invariants is constructed using approximate, 
two-flavor, analytic solutions for RGEs. These invariants exhibit explicitly the 
correlation between quark flavor mixings and mass ratios from electroweak scale to GUT scale in the context of the 
SM, DHM and MSSM models. The well known empirical 
relations $\theta _{23}\propto m_s /m_b $, $\theta _{13}\propto m_d /m_b$ at electroweak scale can thus be understood as the result of 
renormalization evolution toward the infrared point. The validity of this approximation 
is evaluated by comparing the numerical solutions with the analytical approach. 
It is found that the scale dependence of these quantities for general three flavor 
mixing follows closely these invariants up to the GUT scale.

\vspace{60pt}

\pagebreak

\setcounter{page}{1}
\setcounter{footnote}{0}
\renewcommand{\thefootnote}{\arabic{footnote}}
\vspace{15pt}
\begin{flushleft}
{\Large I. Introduction}
\end{flushleft}
\vspace{15pt}


The mixing of quarks arises from the mismatch between quark mass eigenstates and those that participate 
in the weak interactions. The transformation from the weak eigenstates to the mass eigenstates are 
represented by the CKM matrix, which contains three mixing angles and a complex phase, giving 
rise to the phenomena of CP violation. 

   A full knowledge of the CKM matrix is therefore crucial in accounting for observed data and in 
constructing theories beyond the standard model. Since the CKM is obtained from the diagonalization 
of the mass matrices, it is perhaps not surprising that the flavor mixing parameters are correlated to the quark masses. 

   It is well known that both the observed quark mass spectrum and the flavor mixing parameters 
exhibit a strong hierarchical structure. 
The input quark masses at $M_W$ scale are roughly taken in their central values as follows [1](Precisely, as for the quark masses, one need consider the QCD effect when the energy is below $M_W$. However, in the present context, this is not much relevant for us to draw our main conclusion later on) 
$$
m_u\approx 2.4MeV, m_c\approx 1.27GeV, m_t\approx 171GeV
\\
$$                                         
$$
m_d\approx 4.75MeV, m_s\approx 0.104GeV, m_b\approx 4.2GeV                       
\eqno{(1.1)}
$$
They satisfy approximately the following relations [2, 3, 4]  
$$
m_u:m_c:m_t\propto \lambda ^8 :\lambda ^4 :1
\\
$$                                                                      
$$
m_d:m_s:m_b\propto \lambda ^4 :\lambda ^2 :1
\eqno{(1.2)}
$$                                                                                                                 
where $\lambda  = 0.22$. 

   There exists a number of parameterizations of the CKM matrix. They include the Wolfenstein 
parameterization [5], as well as the  standard parameterization  [1], 
$$
V = \left( {\begin{array}{*{20}c}
   {c_{12} c_{13} } & {s_{12} c_{13} } & {s_{13} e^{ - i\delta _{13} } }  \\
   { - s_{12} c_{23}  - c_{12} s_{23} s_{13} e^{i\delta _{13} } } & {c_{12} c_{23}  - s_{12} s_{23} s_{13} e^{i\delta _{13} } } & {s_{23} c_{13} }  \\
   {s_{12} s_{23}  - c_{12} c_{23} s_{13} e^{i\delta _{13} } } & { - c_{12} s_{23}  - s_{12} c_{23} s_{13} e^{i\delta _{13} } } & {c_{23} c_{13} }  \\
\end{array}} \right) 
\eqno{(1.3)}
$$          
where
$$
\theta _{13}  \propto V_{ub}  \propto \lambda ^4, \theta _{23}  \propto V_{cb}  \propto \lambda ^2. 
\eqno{(1.4)}
$$                                                                                                   
Obviously, both the quark masses and the CKM matrix show the hierarchical structure with the 
parameter $\lambda  = 0.22$, and the CKM mixing angles and the quark mass ratios have relations 
$$
\theta _{13}  \propto \frac{{m_d }}{{m_b }}, \theta _{23}  \propto \frac{{m_s }}{{m_b }}  
\eqno{(1.5)}
$$                                                           
at the  $M_W$ scale.

   On the other hand, all physical observables remain unchanged under
the transformation of the redefinition of the phases of the quark fields, i.e., the CKM matrix can be 
rephrased as follows       
$$           
V \to V' = PVQ
\eqno{(1.6)}  
$$                                                                                                          
without changing any physics, where $P$ and $Q$ are diagonal phase matrices. 

   In particle physics one of the substantial problems is that of explaining the quark masses and their mixings. In the quark sector, both the mass ratios and mixing parameters exhibit rather large hierarchies. This observed pattern of fermion masses and mixings does not look accidental. In regards to the relations (1.5), when one tries to sort out possible clues contained therein, we need also keep in mind that these parameters are all measured at low energies. Since the mass matrix evolves with energy, it is necessary to bring renormalization effects into the picture. Therefore, this relation can be extrapolated to all energies with the aid of 
renormalization group equations. As the energy scale changes, one expects the pattern of regularity to evolve as well.
 
  The problem of the RGE of the quark Yukawa couplings has been studied in many papers [6-12]. However, as for the relation Eq.(1.5), one might be tempted to ask 
  
(1) What is the origin of the empirical correlation Eq.(1.5), especially in theory? 

(2) Then what is the correlation between quark flavor mixings and mass ratios in all energy scale?

A complete theory in this regard is certainly lacking and need to be further explored. It is the purpose of the paper, without giving a complete rigorous analysis of the whole parameter space, to untangle these intriguing questions by using the renormalization equations. We hope that our research work could shed some light on the questions of quark masses and mixings.   

   On the other hand, much research work has been focused on the infrared (quasi)fixed points of the renormalization group equations[13-15]. It has been found that, at these (quasi)fixed points, the quark masses have a large hierarchical structure, and the mixing angles vanish. However, in their neighborhood, small angles will be generated through renormalization effects. In this paper, it turns out that, under certain simplifying assumptions, one obtains a set of RGE invariants. At low energy scale, when these fixed points are approached, due to the existence of these RGE invariants, their low energy limits will naturally lead to the correlations described by Eq.(1.5). Thus, this suggests that relations between mass ratios and mixing angles are dynamical in origin. To assess the accuracy of the simplifying assumptions, 
we compare our results numerically with the full-fledged three flavor RGE. It is found that, for a range of parameters, the two methods yield very similar results.
  
  In this paper we make use of the parameterization of CKM matrix proposed in Ref.[16,17]. There, a set of parameters of the CKM matrix was introduced, which is independent of the phase matrices $P$ and $Q$ given by Eq.(1.6). In terms of these rephrasing invariant parameters, it turns out that the RGE are also simpler than those given in terms of other parameters. They are also amenable to a general analysis and enable us to draw our conclusions easily in what follows. 

   The organization of the paper is as follows. In section II, we start with the RGE of the rephrasing 
invariant parameters of the CKM matrix. A set of RGE invariants is then constructed from the 
approximate analytical solutions assuming simple patterns of the CKM matrix. We find, under the 
infrared approximation, they lead to small mixing angles and large mass ratios. Therefore, it shows 
that, the empirical relations for the mixing angles $\theta _{13}$ and $\theta _{23}$ have an origin from the renormalization 
group invariants. The validity and precision of these correlation relations are also evaluated by numerical calculations in the presence of three flavor mixing with respect to the standard model(SM), the double Higgs model(DHM) and the minimal supersymmetric standard model(MSSM) of electroweak interactions from low energy up to the high energy scale. It is found that the scale dependence of these quantities for general three flavoring mixing follows closely these invariants, with a correction term of the order of $O(\lambda ^6 )$ up to the GUT scale. In addition, in 
order to evaluate the approximation condition for these RGE invariants, their renormalization flows 
as functions of the input mass ratios are also plotted. Finally, in subsection 2.4, we present graphically 
the RGE flow of these quantities running from superhigh energy scales to the weak scale, as illustrated 
in the framework of a maximal 
predictability model [18]. And section III is devoted to discussions and summaries. The appendix is 
devoted to exhibiting the details of the RGE used in this paper.

\vspace{15pt}
\begin{flushleft}
{\Large II. RGE Invariants and Quark Flavor Mixings}
\end{flushleft}
\vspace{5pt}
\begin{flushleft}
{\Large 2.1 Rephrasing Invariant Parameters}
\end{flushleft}
\vspace{15pt}

    There are many possible ways to parameterize the CKM matrix. However, since it can be 
multiplied by rephrasing matrices without changing its physical contents, we will choose to use 
a set which is manifestly rephrasing invariant [16]. This set turns out to exhibit the hierarchical 
feature clearly. In addition, the resulting RGE are simpler than those given in terms of other 
parameters and are amenable to a general analysis. These parameters are related to $|V_{ij} |^2$ by 
$$             
W = \left( {\begin{array}{*{20}c}
   {|V_{11} |^2 } & {|V_{12} |^2 } & {|V_{13} |^2 }  \\
   {|V_{21} |^2 } & {|V_{22} |^2 } & {|V_{23} |^2 }  \\
   {|V_{31} |^2 } & {|V_{32} |^2 } & {|V_{33} |^2 }  \\
\end{array}} \right) \\ 
  = \left( {\begin{array}{*{20}c}
   {x_1  - y_1 } & {x_2  - y_2 } & {x_3  - y_3 }  \\
   {x_3  - y_2 } & {x_1  - y_3 } & {x_2  - y_1 }  \\
   {x_2  - y_3 } & {x_3  - y_1 } & {x_1  - y_2 }  \\
\end{array}} \right) \\  
\eqno{(2.1.1)}                                                                                                          
$$                             
Also, these $(x,y)$ parameters satisfy the constraints 
$$
(x_1  + x_2  + x_3 ) - (y_1  + y_2  + y_3 ) = 1 
\\
$$
$$                                                                                
x_1 x_2  + x_2 x_3  + x_3 x_1  = y_1 y_2  + y_2 y_3  + y_3 y_1
\eqno{(2.1.2)}                                                                                                          
$$                                                                                                   
which can be derived form the unitarity condition of the CKM matrix with $\det V = 1$. These constraints 
leave us with four independent parameters, which is consistent with the standard parameterization Eq.(1.3) 
with three mixing angles and a phase. As a result, all of the measurable quantities $\left| {V_{ij} } \right|^2$ are directly related to 
the $(x,y)$ variables. These parameters are found to have an explicit hierarchy structure [17]
$$
x_1  = O(1), x_2  = O(\lambda ^6 ), x_3  = O(\lambda ^6 )  
\\
$$                              
$$                                                                                
y_1  = O(\lambda ^4 ),y_2  = O(\lambda ^2 ). y_3  = O(\lambda ^8 ) 
\\ \eqno{(2.1.3)}                                                                                                          
$$                                                                                                                                                                                                    
The evolution equations of the eigenvalues of the Yukawa coupling matrices are [6-12]
$$
16{\pi ^2}\frac{{dh_i^2}}{{dt}} = h_i^2[{a_d} + bh_i^2 + 2c\sum\limits_j {f_j^2} |{V_{ji}}{|^2}] \\ 
\eqno{(2.1.4)}                                                                                                          
$$
and
$$
16{\pi ^2}\frac{{df_i^2}}{{dt}} = f_i^2[{a_u} + bf_i^2 + 2c\sum\limits_j {h_j^2} |{V_{ij}}{|^2}] \\ 
\eqno{(2.1.5)}                                                                                                          
$$
Here, $t = \ln (\mu /M_W )$, where $\mu$ is the energy scale and $M_W$ stands for the $W$ boson mass, i.e. we choose $M_W$ as the renormalization point, and use quark masses and mixing angles at $M_W$ to fix the initial conditions. Also, $f_i$ and $h_i$ denote the eigenvalues of the Yukawa coupling matrices 
for the up-type and down-type quarks, respectively. Finally, for the coefficients $a_u,a_d,b,c$, we take the notation in Ref.[12]. Specifically, the constant $c$ depends on the model used to run the RGE, and $c =  - 3/2,1/2, 1$ for SM, DHM, and MSSM respectively. Furthermore, the running of the CKM matrix elements  
has the explicit form
$$
16{\pi ^2}\frac{{d{V_{ij}}}}{{dt}} = c[\sum\limits_{l,k \ne i} {{F_{ik}}} h_l^2{V_{il}}V_{kl}^*{V_{kj}} + \sum\limits_{m,k \ne j} {{H_{jk}}} f_m^2V_{mk}^*{V_{mj}}{V_{ik}}] \\ 
\eqno{(2.1.6)}                                                                                                          
$$
where $F_{ik}  = \frac{{f_i^2  + f_k^2 }}{{f_i^2  - f_k^2 }},
 H_{jk}  = \frac{{h_j^2  + h_k^2 }}{{h_j^2  - h_k^2 }}$. 
  
   In terms of the $(x,y)$ parameters, these equations can be reformulated as follow
$$
- 16\pi ^2 \frac{{dx_i }}{{dt}} = c\{ (\Delta f_{23} ,\Delta f_{31} ,\Delta f_{12} )A_i \left( {\begin{array}{*{20}c}
   {H_{23} }  \\
   {H_{31} }  \\
   {H_{12} }  \\
\end{array}} \right) \\ 
  + (\Delta h_{23} ,\Delta h_{31} ,\Delta h_{12} )B_i \left( {\begin{array}{*{20}c}
   {F_{23} }  \\
   {F_{31} }  \\
   {F_{12} }  \\
\end{array}} \right)\}  \\ 
$$       
$$
- 16\pi ^2 \frac{{dy_i }}{{dt}} = c\{ (\Delta f_{23} ,\Delta f_{31} ,\Delta f_{12} )A'_i \left( {\begin{array}{*{20}c}
   {H_{23} }  \\
   {H_{31} }  \\
   {H_{12} }  \\
\end{array}} \right) \\ 
  + (\Delta h_{23} ,\Delta h_{31} ,\Delta h_{12} )B'_i \left( {\begin{array}{*{20}c}
   {F_{23} }  \\
   {F_{31} }  \\
   {F_{12} }  \\
\end{array}} \right)\}  
\\ \eqno{(2.1.7)}                                                                                                          
$$          
in which $\Delta f_{ij}  = f_i^2  - f_j^2, \Delta h_{ij}  = h_i^2  - h_j^2,$ and $A_i(A'_i )$ and $B_i(B'_i )$ are summarized in the appendix.
The scaling dependence of the eigenvalue ratios of the Yukawa couplings can be further written as   
$$             
16\pi ^2 \frac{d}{{dt}}\left( {\begin{array}{*{20}c}
   {\ln r_{23} }  \\
   {\ln r_{31} }  \\
   {\ln r_{12} }  \\
\end{array}} \right) = \frac{b}{2}\left( {\begin{array}{*{20}c}
   {\Delta h_{23} }  \\
   {\Delta h_{31} }  \\
   {\Delta h_{12} }  \\
\end{array}} \right) \\ 
  + c\left( {\begin{array}{*{20}c}
   {x_1  + y_1 } & {x_3  + y_2 } & {x_2  + y_3 }  \\
   {x_2  + y_2 } & {x_1  + y_3 } & {x_3  + y_1 }  \\
   {x_3  + y_3 } & {x_2  + y_1 } & {x_1  + y_2 }  \\
\end{array}} \right)\left( {\begin{array}{*{20}c}
   {\Delta f_{23} }  \\
   {\Delta f_{31} }  \\
   {\Delta f_{12} }  \\
\end{array}} \right)
\\ \eqno{(2.1.8)}                                                                                                            
$$                         
where $r_{ij}  = h_i /h_j$, and $b = 3,3,6$ for SM, DHM, and MSSM respectively. For our purposes, 
it is convenient to reformulate these equations in terms of the $sinh$ function,                                                                    
\begin{align*}
&16\pi ^2 \frac{d}{{dt}}\left( {\begin{array}{*{20}c}
   {\ln [\sinh (\ln r_{23} )]}  \\
   {\ln [\sinh (\ln r_{31} )]}  \\
   {\ln [\sinh (\ln r_{12} )]}  \\
\end{array}} \right) \\ 
= &\frac{b}{2}\left( {\begin{array}{*{20}c}
   {H_{23} \Delta h_{23} }  \\
   {H_{31} \Delta h_{31} }  \\
   {H_{12} \Delta h_{12} }  \\
\end{array}} \right) 
  + c\left( {\begin{array}{*{20}c}
   {H_{23} (x_1  + y_1 )} & {H_{23} (x_3  + y_2 )} & {H_{23} (x_2  + y_3 )}  \\
   {H_{31} (x_2  + y_2 )} & {H_{31} (x_1  + y_3 )} & {H_{31} (x_3  + y_1 )}  \\
   {H_{12} (x_3  + y_3 )} & {H_{12} (x_2  + y_1 )} & {H_{12} (x_1  + y_2 )}  \\
\end{array}} \right)\left( {\begin{array}{*{20}c}
   {\Delta f_{23} }  \\
   {\Delta f_{31} }  \\
   {\Delta f_{12} }  \\
\end{array}} \right)                       
\tag{2.1.9}
\end{align*} 
The relation Eq.(1.5) can be extrapolated from the weak scale up to the scale $M_X$ by means of 
these renormalization group equations. Using the hierarchy of the Yukawa coupling matrices for 
the up and down type quarks, i.e., $y_u ^2  \approx O(1),y_d ^2  \approx O(\lambda ^5 )$[9, 10], 
as well as the hierarchy of 
the $x_i ,y_i$ parameters, Eq.(2.1.7) can be approximated by the following equations 
$$
- 16\pi ^2 \frac{{dx_i }}{{dt}} = c(\Delta f_{23} ,\Delta f_{31} ,\Delta f_{12} )A_i \left( {\begin{array}{*{20}c}
   {H_{23} }  \\
   {H_{31} }  \\
   {H_{12} }  \\
\end{array}} \right) \\,
$$
$$ 
  - 16\pi ^2 \frac{{dy_i }}{{dt}} = c(\Delta f_{23} ,\Delta f_{31} ,\Delta f_{12} )A'_i \left( {\begin{array}{*{20}c}
   {H_{23} }  \\
   {H_{31} }  \\
   {H_{12} }  \\
\end{array}} \right) \\ 
\eqno{(2.1.10)}                                                                                                            
$$                         
Likewise, Eq.(2.1.9) can be written as   
$$
16\pi ^2 \frac{d}{{dt}}\left( {\begin{array}{*{20}c}
   {\ln [\sinh (\ln r_{23} )]}  \\
   {\ln [\sinh (\ln r_{31} )]}  \\
   {\ln [\sinh (\ln r_{12} )]}  \\
\end{array}} \right) \\ 
 =c\left( {\begin{array}{*{20}c}
   {H_{23} (x_1  + y_1 )} & {H_{23} (x_3  + y_2 )} & {H_{23} (x_2  + y_3 )}  \\
   {H_{31} (x_2  + y_2 )} & {H_{31} (x_1  + y_3 )} & {H_{31} (x_3  + y_1 )}  \\
   {H_{12} (x_3  + y_3 )} & {H_{12} (x_2  + y_1 )} & {H_{12} (x_1  + y_2 )}  \\
\end{array}} \right)\left( {\begin{array}{*{20}c}
   {\Delta f_{23} }  \\
   {\Delta f_{31} }  \\
   {\Delta f_{12} }  \\
\end{array}} \right) \\ 
\eqno{(2.1.11)}                                                                                                            
$$
We will now turn to solving these approximate RGE. 

\vspace{15pt}
\begin{flushleft}
{\Large 2.2 RGE Invariants}
\end{flushleft}
\vspace{15pt}

    There have been many phenomenological attempts to relate the quark mixing parameters to the values 
of the quark masses. It would be more appealing if the correlation is stable against the RG evolution up the 
high energy scale. Clearly, it would be nicer to have the RGE for those observables in analytic form when one 
studies the scale dependence of the CKM elements and the quark mass ratios. In fact, as pointed out in 
Ref.[19], in the neutrino sector, a set of RGE invariants is found which bridges the neutrino Yukawa couplings 
and  mixings. There, it is found that the infrared fixed point corresponds to the small neutrino mixing angles 
and infinite mass hierarchies. Based on the same footing of quarks and leptons for some GUT theories, it motivates us that there might be a plausible scenario in the quark sector as well; one would have a natural explanation for large mass hierarchies and small CKM mixing angles, if it results from the RGE evolutions as it approaches the infrared fixed points. 

   We first consider the correlation associated with (2.1.10) and (2.1.11) for the simple 
case of two flavor mixings in (2.1.1). Their analyses are certainly less demanding when 
the flavor mixing involves the second and third family only, i.e., the structure of the 
CKM matrix has the form
$$
W =  \\ 
 \left( {\begin{array}{*{20}c}
   1 & 0 & 0  \\
   0 & {x_1  - y_2 } & {x_2  - y_1 }  \\
   0 & {x_2  - y_1 } & {x_1  - y_2 }  \\
\end{array}} \right) \\ 
  = \left( {\begin{array}{*{20}c}
   1 & 0 & 0  \\
   0 & {\cos ^2 \theta _{23} } & {\sin ^2 \theta _{23} }  \\
   0 & {\sin ^2 \theta _{23} } & {\cos ^2 \theta _{23} }  \\
\end{array}} \right) \\ 
\eqno{(2.2.1)}                                                                                                            
$$
and this can be constructed by setting $x_2  - y_2  = 0,x_3  - y_3  = 0, x_3  - y_2  = 0$, and $x_2  - y_3  = 0$ in Eq.(2.1.1). On the other hand, in derivation of Eq.(2.1.10) and (2.1.11), we assume there is a very large mass hierarchy between the up and down type quarks. As a result, as for the pure mixing between the first two family quarks, we do not expect Eqs.(2.1.10, 2.1.11) are good approximations in the first place. 

   From Eq.(2.1.10) and (2.1.11), we can find the following renormalization correlation 
relations 
$$
\frac{1}{{x_1  - y_2 }}\frac{d}{{dt}}(x_1  - y_2 ) + \frac{1}{{x_2  - y_1 }}\frac{d}{{dt}}(x_2  - y_1 ) + \frac{d}{{dt}}\ln [\sinh (\ln r_{23} )]^2  = 0 
\eqno{(2.2.2)}                                                                                                            
$$
or,
$$
\frac{d}{{dt}}\ln [(x_1  - y_2 )(x_2  - y_1 )\sinh ^2 [\ln \frac{{h_3 }}{{h_2 }}])] = 0 
\eqno{(2.2.3)}  
$$                                                                                                                                                      
Hence, we have the RGE invariant quantity
$$
\Re _{23}  = 2[(x_1  - y_2 )(x_2  - y_1 )]^{1/2} \sinh [\ln \frac{{h_3 }}{{h_2 }}] = const 
\eqno{(2.2.4)}
$$                                                                                                                                                        
or, in terms of the mixing angle given by Eq.(2.2.1), it can be written as
$$
\frac{d}{{dt}}\ln (\frac{1}{2} \sin (2\theta _{23} )\sinh [\ln \frac{{h_3 }}{{h_2 }}]) = 0  
\eqno{(2.2.5)}                                                                                                                                                       
$$
Therefore, for the special structure (2.2.1), the renormalization group invariant has the form:
$$
\Re _{23}  =\sin (2\theta _{23} )\sinh [\ln \frac{{h_3 }}{{h_2 }}] = const  
\eqno{(2.2.6)}                                                                                                                                                       
$$                                                                          
Obviously, this RGE invariant, which is an approximate analytic solution resulting from Eqs.(2.1.7) 
and (2.1.9) for two flavor mixing, is invariant for the energy evolution from the EW scale up to the 
GUT scale for the SM, DHM and MSSM respectively, and the value of the proportionality constant is determined by the physics at 
certain energy scale. 

  In the hierarchical limit, i.e., when $h_3 \gg h_2$,we have 
$$
 \sinh [\ln \frac{{h_3 }}{{h_2 }}] \approx \frac{1}{2}\frac{{h_3 }}{{h_2 }}\eqno{(2.2.7)}                              $$                                                                                                                                                   
Moreover, at the low energy limit, it is well known that there exists (quasi)fixed points for the RGE of the Yukawa coupling matrices [13-15]. At these points, it is found that the quarks have zero mixing angles. But in their neighborhood, small angles will be generated through renormalization effects. From Eq.(2.1.7), for example, we can easily observe that there is a fixed point at $x_1  = 1, x_i  = y_i  = 0, i \ne 1$ for the zero mixing angles. As a result, when these infrared fixed points are approached under the low energy approximation, according to Eq.(2.2.6), the small mixing angle will naturally lead to the large quark mass hierarchical structure                                                            
$$
\sinh [\ln \frac{{h_3 }}{{h_2 }}]\propto \frac{{1 }}{{\theta _{23}}}  
\eqno{(2.2.8)}                                                                                                                                                       
$$                                                                              
Alternatively, it follows that, by assuming the proportionality constant is order $1$ at certain energy scale and considering Eqs.(2.2.6, 2.2.7), in the neighborhood of these (quasi)fixed points we have
$$
\theta _{23}  \approx \frac{{m_s }}{{m_b }} \propto \lambda ^2  
\eqno{(2.2.9)}                                                                                                                                                       
$$                                                                              
The accuracy of these relations can be evaluated by studying the case of general three family 
mixings in Eq.(2.1.1), i.e., $x_2  - y_2, x_3  - y_3, x_3  - y_2,$ and  
$x_2  - y_3 $, are non-vanishing and their 
contributions in the RG evolution of $\Re _{23}$ are considered. $\Re _{23}$ given by Eq.(2.2.6) becomes scale 
dependent and we write
$$
\Re _{23}  = 2[(x_1  - y_2 )(x_2  - y_1 )]^{1/2} \sinh [\ln \frac{{h_3 }}{{h_2 }}]|_{3 \times 3}  
=const+ correction \hspace{3pt} terms             
\eqno{(2.2.10)}                                                                                                                                                       
$$                                                                              
Or, approximately, in this three family mixing case, we can get
$$
\sin (2\theta _{23} )\sinh [\ln \frac{{h_3 }}{{h_2 }}] = const+ correction \hspace{3pt} terms             
\eqno{(2.2.11)}                                                                                                                                                       
$$                                                                              
We then evaluate $\Re _{23}$ numerically. In Figure 1, we plot the scale dependence of $\Re _{23}$ for the general three flavor mixings for the SM, DHM, and MSSM, respectively. Here, for definiteness, we use the proposed values of the quark masses in (1.1) and the initial values of 
CKM matrix elements in Ref.[1] at $t = 0$ as an exemplary example. Then the RGE invariant $\Re _{23}$ is given by $\Re _{23} |_{t = 0}  = 2[(x_1  - y_2 )(x_2  - y_1 )]^{1/2} \sinh [\ln \frac{{h_3 }}{{h_2 }}]|_{2 \times 2} =  2[(x_1  - y_2 )(x_2  - y_1 )]^{1/2} \sinh [\ln \frac{{h_3 }}{{h_2 }}]|_{3 \times 3} = 1.67 $. 
Now, $\Re _{23}$ evolves as a function 
of the momentum from the weak scale all the way up to the grand unification scale. For these three 
models, the numerical correction of the first family to the exact analytical solution is explicitly 
illustrated through the evolution trajectories in the graph.  

    We find that the variations of $\Re _{23}$ are rather slow for each of the models. Specifically, in the 
SM, although its variation is relatively fast, the first family interference effects are not sizable 
and $\Re _{23}$ changes at a level of $0.01\%$ between $M_W$ and the large $10^{15}$ Gev scale.  For the MSSM 
and the DHM, the derivations from $\Re _{23} =const$ are even less compared to those for the SM. 
As a result, it is found that the evolution of $\Re _{23}$ closely follows the RGE invariant $\Re _{23}$ given 
by the two flavor mixing case. The correlation of flavor mixing angles and the mass ratios 
can be approximated as the following
$$
\sin (2\theta _{23} )\sinh [\ln \frac{{h_3 }}{{h_2 }}] = O(1) + O(\lambda ^6 )
\eqno{(2.2.12)}                                                                                                                                                       
$$                                                                              
with a correction term of the order of $O(\lambda ^6 )$ up to the GUT scale. 
\begin{figure}
\centering
\includegraphics[width=6cm,height=5cm]{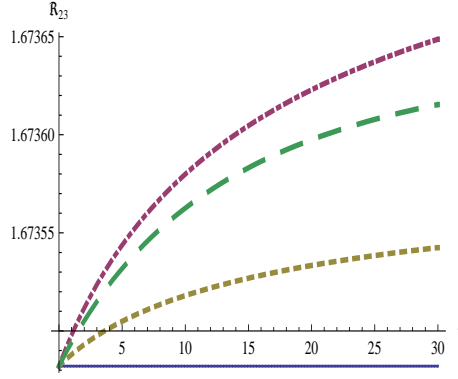}
\caption{Renormalization evolution of the quantity $\Re _{23}$. The top dotdashed line stands 
for the SM evolution. The dashed and dotted lines are for the MSSM and 
DHM, respectively. The flat solid line is the RGE invariant for two 
flavor mixing.}\label{fig:rgeinv23.eps} 
\end{figure}

     Likewise, consider the mixing of the first and third families only by setting
$x_2  - y_2  = 0,x_3  - y_2  = 0,x_2  - y_1  = 0,$ and $x_3  - y_1  = 0$ in Eq.(2.1.1). 
The structure of the CKM matrix has the form
$$
 W = \left( {\begin{array}{*{20}c}
   {x_1  - y_2 } & 0 & {x_3  - y_3 }  \\
   0 & 1 & 0  \\
   {x_3  - y_3 } & 0 & {x_1  - y_2 }  \\
\end{array}} \right) \\ 
  = \left( {\begin{array}{*{20}c}
   {\cos ^2 \theta _{13} } & 0 & {\sin ^2 \theta _{13} }  \\
   0 & 1 & 0  \\
   {\sin ^2 \theta _{13} } & 0 & {\cos ^2 \theta _{13} }  \\
\end{array}} \right) 
\eqno{(2.2.13)}                                                                                                                                                       
$$                                                                              
Applying Eq.(2.1.10) and (2.1.11) leads to the following renormalization differentiation 
equation
$$
\frac{1}{{x_1  - y_2 }}\frac{d}{{dt}}(x_1  - y_2 ) + \frac{1}{{x_3  - y_3 }}\frac{d}{{dt}}(x_3  - y_3 ) + \frac{d}{{dt}}\ln [\sinh (\ln r_{31} )]^2  = 0 
\eqno{(2.2.14)}                                                                                                                                                       
$$                                                                              
Namely, 
$$
\frac{d}{{dt}}\ln [(x_1  - y_2 )(x_3  - y_3 )\sinh ^2 [\ln \frac{{h_3 }}{{h_1 }}])] = 0 
\eqno{(2.2.15)}                                                                                                                                                       
$$         
or
$$
\frac{d}{{dt}}\ln (\frac{1}{2}\sin (2\theta _{13} )\sinh [\ln \frac{{h_3 }}{{h_1 }}]) = 0 
\eqno{(2.2.16)}                                                                                                                                                       
$$         
Thus we have the following RGE invariant corresponding to the exact analytic 
solution of Eqs.(2.1.10) and (2.1.11) for the special pattern (2.2.13)
$$
\Re _{13}  = \sin (2\theta _{13} )\sinh [\ln \frac{{h_3 }}{{h_1 }}] = const 
\eqno{(2.2.17)}                                                                                                                                                       
$$         
Again, the proportionality constant is determined by the physics at certain energy scale. And at the low energy limit, we have a small mixing angle near the fixed points due to renormalization effect. There, by further assuming the proportionality constant is order $1$ in Eq.(2.2.17), that gives us 
$$
\theta _{13}  \approx \frac{{h_1 }}{{h_3 }} \propto \frac{{m_d }}{{m_b }} \propto \lambda ^4  
\eqno{(2.2.18)}                                                                                                                                                       
$$         

    In the general three flavor mixing case, the evolution of $\Re _{13}$ becomes energy 
dependent and is evaluated numerically. In Figure 2, we plot the renormalization group 
evolution of $\Re _{13}$ for the SM, DHM, and MSSM respectively. 
\begin{figure}
\centering
\includegraphics[width=6cm,height=5cm]{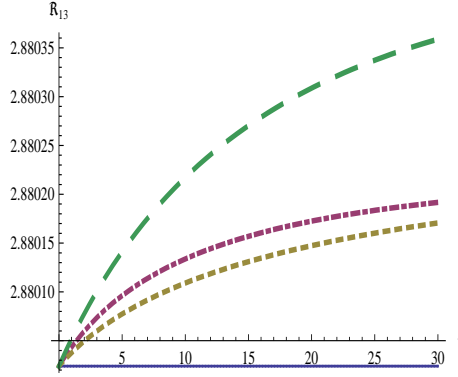}
\caption{Renormalization evolution of the quantity $\Re _{13}$, 
using the same notation as is Fig.1.}\label{fig:rgeinv13.eps} 
\end{figure}
                        
    From Figure 2, one finds that the difference between the analytic solution and the 
full MSSM is significant compared to other models. But the relative deviations for $\Re _{13}$ 
still less than $0.02\%$ in the whole range of $t$ up to the GUT scale, and the DHM is special 
in that it allows a smallest deviation in the presence of three flavor mixings. One can conclude that 
$$
\Re _{13}  = 2[(x_1  - y_2 )(x_3  - y_3 )]^{1/2} \sinh [\ln \frac{{h_3 }}{{h_1 }}] = O(1) + O(\lambda ^6 ) 
\eqno{(2.2.19)}                                                                                                                                                       
$$         
where the constant, as an exemplary example, has been fixed to be 2.88 by considering the initial values of the 
current quark masses and mixing angles at electroweak scale. All in all, $\Re _{13}$ is practically 
constant even up to the GUT scale. The observed symmetry (2.2.20) arises as a result 
of infrared fixed points of the low energy theory. In the neighborhood of the fixed point, 
the mixing angles attain small values and the mass hierarchy becomes large. Irrespective 
of the specific models, the evolution of $\Re _{13}$ has interesting infrared fixed points. The 
behavior of the mass and mixing hierarchy can thus be explained from the point of 
view of renormalization group.            
       
\vspace{15pt}
\begin{flushleft}
{\Large 2.3 Mass Hierarchy Dependence}
\end{flushleft}
\vspace{15pt}

   As described above, a set of renormalization invariants for all energy scales can be 
constructed from Eqs.(2.1.10) and (2.1.11) for special patterns of the CKM matrix. 
They are good approximations of Eqs.(2.1.7) and (2.1.9) under the condition that the 
up and down type quarks have large hierarchical structure. It is interesting to investigate 
the accuracy of these approximations as one varies the degree of mass hierarchy. For 
this purpose we vary the input mass ratios at low energy, and calculate, at high energy, 
the deviations of $\Re _{ij}$ from constant, as a function of the input. For definiteness, we 
consider such mass dependence for $\Re _{23}$ only, since the behavior of $\Re _{13}$ is similar. 

   In Figure 3, we plot the explicit dependence on the up-down type mass ratios for the 
standard model. At the high energy scale, it is found that 
\begin{align*}
2[(x_1  - y_2 )(x_2  - y_1 )]^{1/2} \sinh [\ln \frac{{h_3 }}{{h_2 }}] = & const + O(\lambda ^4 ) \hspace{3pt} for
\hspace{3pt} m_b /m_t  = 10\%  \\
2[(x_1  - y_2 )(x_2  - y_1 )]^{1/2} \sinh [\ln \frac{{h_3 }}{{h_2 }}] = & const + O(\lambda ^3 ) \hspace{3pt} for 
\hspace{3pt} m_b /m_t  = 20\%  
\tag{2.3.1}
\end{align*} 
Explicitly, the approximation of Eq.(2.2.4) is valid up to the order of $\lambda ^4$ if the mass hierarchy 
is no more than $10\%$. And the deviation from Eq.(2.2.4) will be up to $\lambda ^3$ order when the mass 
ratio is around $20\%$. We are thus assured that Eqs.(2.1.10) and (2.1.11) are good 
approximations if the mass ratio is less than $10\%$. 
\begin{figure}
\centering
\includegraphics[width=6cm,height=5cm]{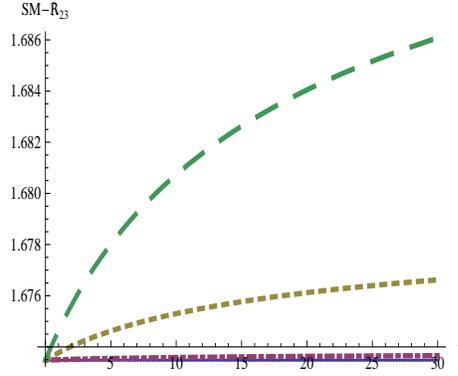}
\caption{Mass hierarchy dependence of the $\Re _{23} $
renormalization evolution of the SM. The dashed line 
is for $m_b /m_t  = 20\%$. The dotted line is when $m_b /m_t  = 10\%$.  
The dotdashed is the normal SM. The solid line is the RGE 
invariant for the two flavoring mixings.}\label{fig:secondgraph3.eps} 
\end{figure}

    Likewise, in Figures 4 and 5, we can find the mass hierarchy dependence of $\Re _{23}$ for 
DHM and MSSM respectively. We use the same notation for the lines as Figure 3. 
\begin{figure}
\centering
\includegraphics[width=6cm,height=5cm]{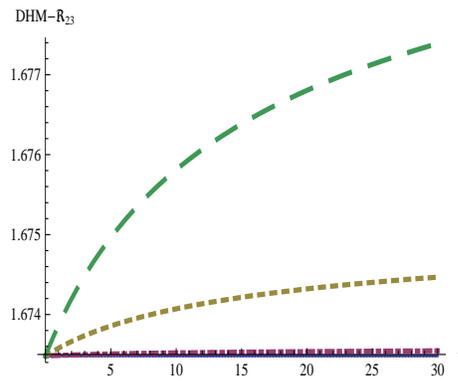}
\caption{Mass hierarchy dependence of $\Re _{23}$
renormalization evolution of DHM.}\label{fig:dhmrge23.eps} 
\end{figure}
Thus, for the DHM, 
\begin{align*}
\Re _{23}  = & const + O(\lambda ^5 ) \hspace{3pt}  for \hspace{3pt}  m_b /m_t  = 10\%  \\ 
\Re _{23}  = & const + O(\lambda ^4 )\hspace{3pt}  for \hspace{3pt} m_b /m_t  = 20\%  
\tag{2.3.2}
\end{align*} 
\begin{figure}
\centering
\includegraphics[width=6cm,height=5cm]{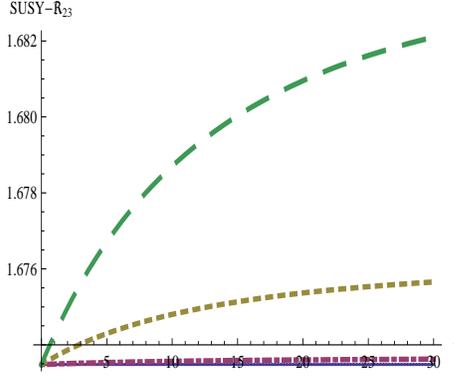}
\caption{Mass hierarchy dependence of $\Re _{23}$
renormalization evolution of MSSM.}\label{fig:susyrge23.eps} 
\end{figure}
As for the MSSM, we have 
$$
\Re _{23}  = const + O(\lambda ^4 )\hspace{3pt}   for \hspace{3pt}  m_b /m_t  = 10\%  
\\ 
$$ 
$$
\Re _{23}  = const + O(\lambda ^3 )\hspace{3pt}   for \hspace{3pt}  m_b /m_t  = 20\% 
\eqno{(2.3.3)}                                                                                                                                                       
$$
Hence, the mass hierarchy dependence of SM and MSSM give us the 
same approximation condition for the validity of Eq.(2.2.4), i.e., they both 
lead to the deviation up to $\lambda ^3$ order when the mass ratio becomes $20\%$. 
However, the DHM is the least sensitive one, where the deviation is not 
appreciable even when the mass ratio reaches $20\%$.   

   Therefore, from the behavior of the evolution of $\Re _{23}$, we can be assured 
that Eq.(2.2.4) is a good approximation for each model if the down-up type 
mass ratio is less than $10\%$, with a correction term of order $\lambda ^4$.  

\vspace{15pt}
\begin{flushleft}
{\Large 2.4 High Energy Scale Evolution}
\end{flushleft}
\vspace{15pt}

    So far, we have studied the scaling behaviors of physical quantities constructed 
from RGE invariants running from low energy to the grand unification scale. However, 
at the GUT scale, the symmetry group of the model may become larger, and this may 
entail additional symmetries or textures of the quark Yukawa coupling matrices. 
Therefore, in order to compare the quark masses and mixings with the present 
experimental data, it is necessary to run the quark masses and mixings from the 
unification scale down to the electroweak scale. Here, it becomes natural and instructive to 
ask the evolution of these physical quantities based on certain symmetry structures from 
GUT scale to low energy scale. A common phenomenological approach is to assume 
special textures for the quark mass matrices and to derive experimentally testable relations 
among quark masses and mixing angles [18].  We focus on the superstring motivated 
pattern of the quark mass matrices proposed in Ref.[18], in which the quark mass matrices 
have a hierarchy structure with a minimal number of free parameters, constrained by the 
underlying symmetry theories. We then analyze how the relations of Eqs.(2.2.12) and (2.2.21) 
evolve from $10^{15}$ Gev to electroweak scale and examine their behaviors which might impact 
low energy physics. The patterns of the mass matrices have the following explicit hierarchy texture
$$
M_u  = a_u \left( {\begin{array}{*{20}c}
   0 & {w^6 \varepsilon ^6 } & 0  \\
   {w^6 \varepsilon ^6 } & {w^4 \varepsilon ^4 } & {w^4 \varepsilon ^4 }  \\
   0 & {w^4 \varepsilon ^4 } & 1  \\
\end{array}} \right), 
 M_d  = a_d \left( {\begin{array}{*{20}c}
   0 & {\varepsilon ^3 e^{ - i\varphi _1 } } & {\varepsilon ^4 e^{ - i\varphi _2 } }  \\
   {\varepsilon ^3 e^{i\varphi _1 } } & {\varepsilon ^2 } & {\varepsilon ^2 e^{ - i\varphi _3 } }  \\
   {\varepsilon ^4 e^{i\varphi _2 } } & {\varepsilon ^2 e^{i\varphi _3 } } & 1  \\
\end{array}} \right) 
\eqno{(2.4.1)}                                                                                                                                                       
$$
After diagonalizing the mass matrixes by unitary transformations, i.e. $U^ +  M_u U = D_u,   
V^ +  M_d V = D_d$,the CKM matrix is given by $V_{CKM}  = U^ +  V$. Correspondingly, the evolutions of 
quantities $\Re _{13}$ and $\Re _{23}$ are plotted as shown in Fig.6 and 7.
\begin{figure}
\centering
\includegraphics[width=6cm,height=5cm]{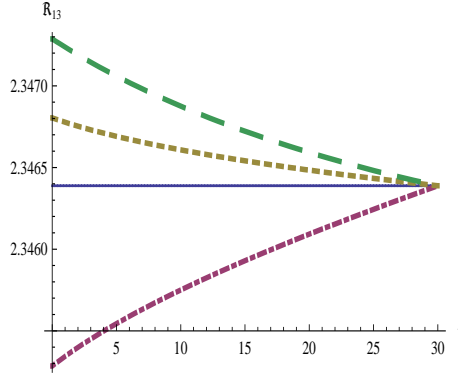}
\caption{Renormalization evolution of $\Re _{13}$}\label{fig:highenergyr13.eps} 
\end{figure}
\begin{figure}
\centering
\includegraphics[width=6cm,height=5cm]{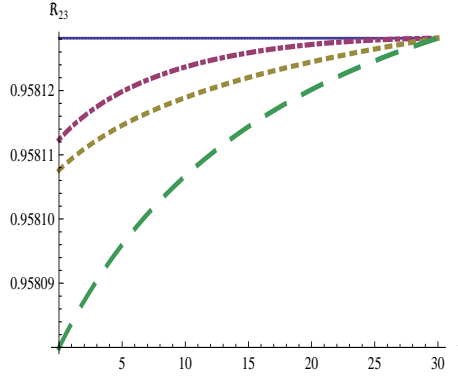}
\caption{Renormalization evolution of $\Re _{23}$. The Solid lines in Figs.6,7 are 
the RGE invariants defined by (2.2.6) and (2.2.19) respectively. The dotdashed 
lines are of the SM. The dotted lines are of the DHM, and the dashed lines are 
for SUSY.  }\label{fig:highenergyr23.eps} 
\end{figure}

For simplicity and definiteness, we have taken the initial input values 
as $a_u  = 120 Gev,a_d  = 0.9 Gev,\varepsilon  = 0.19,w = 1.2$ and $\varphi _1  = 2.2,
\varphi _2  = 2.2 + \pi /2,\varphi _3  = 0$.
The $\Re _{13}$ evolution for the SM has the opposite curvature from the other two models, 
which also has the maximum deviation from the central solid line for RGE invariant. 
It reaches around $0.06\%$ in the low energy limit. As for the evolution of $\Re _{23}$, all the 
models have the same evolution directions for the RGE flows, and the maximal deviation 
occurs for the MSSM, at the level of $O(\lambda ^7 )$. 

   Our results show that, for mass matrices of the form given in Eq.(2.4.1), renormalization 
effects are very small, whether we use the SM, DHM or MSSM. If one intends to improve 
the model predictions at low energies, it is necessary to either change the input values at the 
GUT scale, or one could invoke additional symmetry groups for the renormalization evolution. 
Such extra symmetries could arise from extra dimension theories [20], for instance. In fact, 
the extra Kaluza-Klein states in these theories could change the RGE trajectories substantially. 
We hope to return to this analysis in the future.

\vspace{15pt}
\begin{flushleft}
{\Large III. Conclusion}
\end{flushleft}
\vspace{15pt}

In this paper we find that the general RGE for the quark mass matrices have simple solutions in appropriate two flavor approximations. These solutions can be expressed in the form of energy scale invariants which relate mass ratios and flavor mixing parameters. Specifically, these invariants are 
$$
2[(x_1  - y_2 )(x_3  - y_3 )]^{1/2} \sinh [\ln \frac{{h_3 }}{{h_1 }}] = const \\ 
$$  
$$
2[(x_1  - y_2 )(x_2  - y_1 )]^{1/2} \sinh [\ln \frac{{h_3 }}{{h_2 }}] = const 
\eqno{(3.1)}                                                                                                                                                       
$$   
or 
$$
\sin (2\theta _{13} )\sinh [\ln \frac{{h_3 }}{{h_1 }}] = const \\
$$                        
$$
\sin (2\theta _{23} )\sinh [\ln \frac{{h_3 }}{{h_2 }}] = const  
\eqno{(3.2)}                                                                                                                                                                                                                                                                                                             
$$                                                                           
The validity and precision of these correlation relations are found quite accurate when compared with the complete numerical calculations for general three flavor mixing from low energy up to the GUT energy scale. It is observed that there is little deviation between the two approaches.
From the RGE of Eq.(2.1.7) for the $(x,y)$ variables, as found before, there is a fixed point at $x_1  = 1, x_i  = y_i  = 0, i \ne 1$.  There, physically one can find it will give us zero mixing angles. The physical (quasi)fixed points with zero mixing angles for the Yukawa coupling matrices have been extensive discussed in the past [13-15]. Close to these points, corresponding to small mixing angles, 
Eq.(3.1) implies $\theta _{23}\propto \frac{{m_s }}{{m_b }},  \theta _{13} \propto \frac{{m_d }}{{m_b }}$, 
the well-known empirical relations between 
physical parameters. This suggests a dynamical origin for these empirical relations. And 
the hierarchies in the mass ratios and mixing angles can be the result of renormalization 
evolution as the energy approaches these infrared (quasi)fixed points. 

    Note that, our approximation depends on the assumed large hierarchy between the up and down type quark masses. 
This works well when applied to mixings involving the third family. However, it can not be justified for the mixing between 
the first two light families. Thus, for the Cabibbo angle we do not expect to have a relation analogous to Eq.(3.1). Our 
analysis brings out the qualitative difference between the light and heavy families, in agreement 
with the observed patterns of known physical parameters. \footnote{  
An alternative discussion can be found in Ref.[21]. There, it is shown that the Cabibbo angle and the quark mass ratio have the relation $\tan \theta_c \sqrt{m_s/m_d}=constant$, which is scale invariant from $M_Z$ to the Planck scale.} 

   We have also analyzed the renormalization evolution starting from high energy scales. It is found that, at least for certain existing models, the RGE invariant solutions follow closely the numerical calculations. 
It is our hope that our analysis can provide the impetus for further researches along this direction, so that the success or failure of a model can be better assessed. The fact that the existence of these RGE invariants is rather instructive, especially when one introduces special quark texture at high energy scale. It may also put a substantial constraint on the parameter space of the quark mass matrices, or may serve as important clues in search for new physics symmetry theories. Actually, at high energies, there are many possible choices of the physical parameters, in addition to various options for the 
underlying symmetry groups. A more complete analysis of this problem is beyond the scope of this work, but will be attempted in a future publication.

\vspace{15pt}
\begin{flushleft}
{\Large Acknowledgments
}
\end{flushleft}
\vspace{15pt}

This work is supported in part by NITheP fellowship. The author would 
like to thank Tae-Hun Lee for assistance in computer programming. The author also would like to thank 
T. K. Kuo for his support of this research work.

\pagebreak

\begin{flushleft}
{\Large Appendix
}
\end{flushleft}
\vspace{15pt}
The $A(A')$ and $B(B')$ in Eq.(2.1.7) are defined as follows:
\vspace{15pt}
\vspace{15pt}
\\
$
A_1  = x_1 \left( {\begin{array}{*{20}c}
   {y_1 } & {x_2 } & {x_3 }  \\
   {x_3 } & {y_3 } & {x_2 }  \\
   {x_2 } & {x_3 } & {y_2 }  \\
\end{array}} \right) + \left( {\begin{array}{*{20}c}
   {y_1 x_1 } & {y_3 y_2 } & {y_2 y_3 }  \\
   {y_1 y_2 } & {y_3 x_1 } & {y_2 y_1 }  \\
   {y_1 y_3 } & {y_3 y_1 } & {y_2 x_1 }  \\
\end{array}} \right), 
$
\\
$
A_2  = x_2 \left( {\begin{array}{*{20}c}
   {x_1 } & {y_2 } & {x_3 }  \\
   {x_3 } & {x_1 } & {y_1 }  \\
   {y_3 } & {x_3 } & {x_1 }  \\
\end{array}} \right) + \left( {\begin{array}{*{20}c}
   {y_3 y_1 } & {y_2 x_2 } & {y_1 y_3 }  \\
   {y_3 y_2 } & {y_2 y_3 } & {y_1 x_2 }  \\
   {y_3 x_2 } & {y_2 y_1 } & {y_1 y_2 }  \\
\end{array}} \right),
$
\\
$
 A_3  = x_3 \left( {\begin{array}{*{20}c}
   {x_1 } & {x_2 } & {y_3 }  \\
   {y_2 } & {x_1 } & {x_2 }  \\
   {x_2 } & {y_1 } & {x_1 }  \\
\end{array}} \right) + \left( {\begin{array}{*{20}c}
   {y_2 y_1 } & {y_1 y_2 } & {y_3 x_3 }  \\
   {y_2 x_3 } & {y_1 y_3 } & {y_3 y_1 }  \\
   {y_2 y_3 } & {y_1 x_3 } & {y_2 y_2 }  \\
\end{array}} \right), $
\\ 
$
 B_1  = x_1 \left( {\begin{array}{*{20}c}
   {y_1 } & {x_3 } & {x_2 }  \\
   {x_2 } & {y_3 } & {x_3 }  \\
   {x_3 } & {x_2 } & {y_2 }  \\
\end{array}} \right) + \left( {\begin{array}{*{20}c}
   {y_1 x_1 } & {y_3 y_2 } & {y_2 y_3 }  \\
   {y_1 y_2 } & {y_3 x_1 } & {y_2 y_1 }  \\
   {y_1 y_3 } & {y_3 y_1 } & {y_2 x_1 }  \\
\end{array}} \right), $
\\ 
$
 B_2  = x_2 \left( {\begin{array}{*{20}c}
   {x_1 } & {x_3 } & {y_3 }  \\
   {y_2 } & {x_1 } & {x_3 }  \\
   {x_3 } & {y_1 } & {x_1 }  \\
\end{array}} \right) + \left( {\begin{array}{*{20}c}
   {y_2 y_1 } & {y_1 y_2 } & {y_3 x_2 }  \\
   {y_2 x_2 } & {y_1 y_3 } & {y_3 y_1 }  \\
   {y_2 y_3 } & {y_1 x_2 } & {y_3 y_2 }  \\
\end{array}} \right), 
$
\\ 
$
 B_3  = x_3 \left( {\begin{array}{*{20}c}
   {x_1 } & {y_2 } & {x_2 }  \\
   {x_2 } & {x_1 } & {y_1 }  \\
   {y_3 } & {x_2 } & {x_1 }  \\
\end{array}} \right) + \left( {\begin{array}{*{20}c}
   {y_3 y_1 } & {y_2 x_3 } & {y_1 y_3 }  \\
   {y_3 y_2 } & {y_2 y_3 } & {y_1 x_3 }  \\
   {y_3 x_3 } & {y_2 y_1 } & {y_1 y_2 }  \\
\end{array}} \right), 
$
\\ 
$
 A'_1  = y_1 \left( {\begin{array}{*{20}c}
   {x_1 } & {y_2 } & {y_3 }  \\
   {y_2 } & {y_3 } & {x_2 }  \\
   {y_3 } & {x_3 } & {y_2 }  \\
\end{array}} \right) + \left( {\begin{array}{*{20}c}
   {x_1 y_1 } & {x_3 x_2 } & {x_2 x_3 }  \\
   {x_1 x_3 } & {x_3 x_1 } & {x_2 y_1 }  \\
   {x_1 x_2 } & {x_3 y_1 } & {x_2 x_1 }  \\
\end{array}} \right), 
$
\\ 
$
 A'_2  = y_2 \left( {\begin{array}{*{20}c}
   {y_1 } & {x_2 } & {y_3 }  \\
   {x_3 } & {y_3 } & {y_1 }  \\
   {y_3 } & {y_1 } & {x_1 }  \\
\end{array}} \right) + \left( {\begin{array}{*{20}c}
   {x_3 x_1 } & {x_2 y_2 } & {x_1 x_3 }  \\
   {x_3 y_2 } & {x_2 x_1 } & {x_1 x_2 }  \\
   {x_3 x_2 } & {x_2 x_3 } & {x_1 y_2 }  \\
\end{array}} \right), 
$
\\
$ 
 A'_3  = y_3 \left( {\begin{array}{*{20}c}
   {y_1 } & {y_2 } & {x_3 }  \\
   {y_2 } & {x_1 } & {y_1 }  \\
   {x_2 } & {y_1 } & {y_2 }  \\
\end{array}} \right) + \left( {\begin{array}{*{20}c}
   {x_2 x_1 } & {x_1 x_2 } & {x_3 y_3 }  \\
   {x_2 x_3 } & {x_1 y_3 } & {x_3 x_2 }  \\
   {x_2 y_3 } & {x_1 x_3 } & {x_3 x_1 }  \\
\end{array}} \right), 
$
\\ 
$
 B'_1  = y_1 \left( {\begin{array}{*{20}c}
   {x_1 } & {y_2 } & {y_3 }  \\
   {y_2 } & {y_3 } & {x_3 }  \\
   {y_3 } & {x_2 } & {y_2 }  \\
\end{array}} \right) + \left( {\begin{array}{*{20}c}
   {x_1 y_1 } & {x_2 x_3 } & {x_3 x_2 }  \\
   {x_1 x_2 } & {x_2 x_1 } & {x_3 y_1 }  \\
   {x_1 x_3 } & {x_2 y_1 } & {x_3 x_1 }  \\
\end{array}} \right), 
$
\\ 
$
 B'_2  = y_2 \left( {\begin{array}{*{20}c}
   {y_1 } & {x_3 } & {y_3 }  \\
   {x_2 } & {y_3 } & {y_1 }  \\
   {y_3 } & {y_1 } & {x_1 }  \\
\end{array}} \right) + \left( {\begin{array}{*{20}c}
   {x_2 x_1 } & {x_3 y_2 } & {x_1 x_2 }  \\
   {x_2 y_2 } & {x_3 x_1 } & {x_1 x_3 }  \\
   {x_2 x_3 } & {x_3 x_2 } & {x_1 y_2 }  \\
\end{array}} \right), 
$
\\ 
$
 B'_3  = y_3 \left( {\begin{array}{*{20}c}
   {y_1 } & {y_2 } & {x_2 }  \\
   {y_2 } & {x_1 } & {y_1 }  \\
   {x_3 } & {y_1 } & {y_2 }  \\
\end{array}} \right) + \left( {\begin{array}{*{20}c}
   {x_3 x_1 } & {x_1 x_3 } & {x_2 y_3 }  \\
   {x_3 x_2 } & {x_1 y_3 } & {x_2 x_3 }  \\
   {x_3 y_3 } & {x_1 x_2 } & {x_2 x_1 }  \\
\end{array}} \right). \\ 
$

\pagebreak

\begin{center}
{\bf REFERENCES}
\end{center}
\begin{description}

\item[[1]] C. Amsler et al, Particle Data Group, PL B667, 1 (2008) 

\item[[2]] Shao-Hsuan Chiu, T.K.Kuo and Guo-Hong Wu, Phys.Rev.D62:053014, 2000 [hep-
      ph/0003224]

\item[[3]] T. K. Kuo, S. W. Mansour and G. H. Wu, Phys.Lett.B467:116-125, 1999 [hep-
     ph/9907521] 

\item[[4]] J. A. Aguilar-Saavedra and M. Masip, Phy. Rev. D 54, 6903, 1996

\item[[5]] L. Wolfenstein, Phys. Rev. Lett. 51, 1945 (1983)

\item[[6]] K.S. Babu, Z.Phys.C35:69, 1987

\item[[7]] K. Sasaki, Z. Phys. C 32, 149, 1986

\item[[8]] M. E. Machacek, M. T. Vaughn, Nucl.Phys.B236:221, 1984

\item[[9]] P. Kielanowski, S. R. Juarez W. and J. G. Mora H, Phys.Lett.B479:181-189, 2000 
     [hep-ph/0002062] 

\item[[10]]S.R. Juarez Wysozka, S.F. Herrera H., P. Kielanowski, G. Mora, Phys. Rev. D66 : 116007, 2002 [hep-ph/0206243] 

\item[[11]] H. Gonzalez, S.Rebeca Juarez Wysozka, P. Kielanowski, G. Lopez Castro,  Phys.  
       Lett. B440:94-100, 1998.

\item[[12]]Shao-Hsuan Chiu, T.K. Kuo, Tae-Hun Lee, C. Xiong, Phys.Rev.D79:013012,2009.

\item[[13]] B. Pendleton, Graham G. Ross, Phys.Lett.B98:291,1981.
 
\item[[14]] C. T. Hill, Phys. Rev. D24:691,1981. 

\item[[15]] J. Bagger, S. Dimopoulos, E. Masso,Phys.Rev.Lett.55:920,1985;
J. Bagger, S. Dimopoulos, E. Masso, Nucl.Phys.B253:397,1985.

\item[[16]] T.K.Kuo and Tae-Hun Lee, Phys.Rev.D71:093011, 2005 [hep-ph/0504062]

\item[[17] T.K.Kuo and Lu-Xin Liu, hep-ph/0511037

\item[[18]] G. Branco, D. Emmanuel-Costa, R. Felipe, Phys. Lett. B483, 87, 2000

\item[[19]] T.K.Kuo, Mod. Phys. Lett. A 17, 2355, 2002

\item[[20]] K. Dienes, E. Dudas, T. Gherghetta, Nucl.Phys.B537:47, 1999

\item[[21]] H. Arason, D.J. Castano, E.J. Piard, P. Ramond, Phys. Rev. D47: 232-240,1993.

\end{description}

\end{document}